\documentclass[epsfig]{emulateapj}
\shorttitle{Spectropolarimetry of SN~2011fe}
\shortauthors{Milne et al.}
\begin{document}

\title{Multi-epoch Spectropolarimetry of SN~2011\lowercase{fe}}

\author{
  Peter A. Milne\altaffilmark{1}, G. Grant Williams\altaffilmark{1,2}, \
Amber Porter\altaffilmark{3}, Paul S. Smith\altaffilmark{1}, \
%  Paul S. Smith\altaffilmark{1}, Amber Porter\altaffilmark{3}, \
  Nathan Smith\altaffilmark{1}, \
Mark D. Leising\altaffilmark{3},  Buell T. Jannuzi\altaffilmark{1},  
\& E.M.  Green\altaffilmark{1}
       }

\altaffiltext{1}{University of Arizona, Steward Observatory, 933 N.\
  Cherry Ave., Tucson, AZ 85721.}
\altaffiltext{2}{MMT Observatory, 933 N.\ Cherry Ave., Tucson, AZ
  85721.}
\altaffiltext{3}{118 Kinard Laboratory, Clemson University, Clemson, SC 29634 }

\begin{abstract}

We present multiple spectropolarimetric observations of the nearby
Type~Ia supernova, SN~2011fe in M101, obtained before, during, and
after the time of maximum apparent visual brightness.  
The excellent time coverage of our spectropolarimetry has allowed 
better monitoring of the evolution of polarization features than is 
typical, which has allowed us new insight into the nature of normal SNe~Ia. 
SN~2011fe
exhibits time-dependent polarization in both the continuum and
strong absorption lines.  At early epochs, red wavelengths exhibit a
degree of continuum polarization of up to 0.4\%, likely indicative of
a mild asymmetry in the electron-scattering photosphere. This behavior 
is more common in sub-luminous SNe~Ia than in normal events, such as 
SN~2011fe. 
The degree of polarization across a collection of 
absorption lines varies dramatically from epoch to
epoch.  During the earliest epoch 
a $\lambda$4600-5000 \AA\ complex of absorption lines shows enhanced 
polarization at a different position angle than the continuum. 
We explore the origin of these features, presenting a few possible 
interpretations, without arriving at a single favored ion. During two epochs 
near maximum, the dominant polarization feature is associated with the  
Si~{\sc ii} $\lambda$6355 \AA\ absorption line. This is common for SNe~Ia, 
but for SN~2011fe the polarization of this feature 
increases after maximum light, whereas for other 
SNe~Ia, that polarization feature was strongest before maximum light. 

\end{abstract}


%%%%%%%%%%%%%%%%%%%%%%%%%%%%%%%%%%%%%%%%%%%%%%%%%%%%%%%%%%%%%%%%%%%%%%%%%%
\section{INTRODUCTION}

Type~Ia supernova (SN~Ia) explosions convey information about the
nucleosynthesis by the thermonuclear destruction of a CO white dwarf
(Iwamoto et al.\ 1999), and they provide a way to measure the
expansion of the universe by using their peak magnitudes as
standardizable candles (Phillips 1993). The unexpected finding that
the expansion of the universe is accelerating (Riess et al.\ 1998;
Perlmutter et al.\ 1999) has focused interest on a better
understanding of the SN~Ia explosion mechanism.  It has long been
recognized that there are variations within the SN~Ia category. More
luminous events rise to peak and decline from peak on a longer
timescale than less luminous events (Branch, Fisher \& Nugent 1993;
Phillips 1993).  The majority of events fall within a ``normal''
grouping, although some cases have been recognized where the
luminosity does not correlate with peak width (e.g., Benetti et al.\
2005; Wang et al.\ 2009; Foley \& Kasen 2011). Asymmetries in the
explosion may hold important clues to the explosion mechanism itself,
as well as to the consequent diversity in observed properties.  

Spectropolarimetry has emerged as a powerful probe of SNe~Ia (see
Livio \& Pringle 2011) and of the intervening interstellar (ISM) or 
circumstellar (CSM) material (Patat et al. 2015; Porter et al. 2016). 
The degree of polarization of the continuum
emission is generally lower for SNe~Ia than for core-collapse events,
but it has been detected at significant levels for a range of SN~Ia
sub-classes (Wang \& Wheeler 2008). 
Polarization at the wavelengths of observed absorption 
lines is particularly interesting as it affords the opportunity to 
study the distribution of specific elements within the ejecta. 
This ``line polarization'' has been observed to change markedly near 
maximum light.  The signature Si~{\sc ii} $\lambda$6355 \AA\ line in 
addition to the Ca~{\sc ii} NIR triplet near 8000 \AA\ has exhibited 
polarization in a number of SNe~Ia including 2001el (Wang et al.2003), 
2002bo (Wang et al. 2007), 1997dt, 2002bf, 2003du (Leonard et al. 2005), 
2004S (Chornock \& Filippenko 2008), 2004dt (Wang et al. 2006, Leonard et al. 2005),
2006X (Patat et al. 2009), 2008fp (Cox \& Patat 2014), 
2012fr (Maund et al. 2013), and 2014J (Patat et al. 2015).  
Likewise, the sub-luminous 1999by (Howell et al. 2001), 
2005ke (Patat et al. 2012), and the super-Chandrasekhar explosion 
2009dc (Tanaka et al. 2010) have shown variable polarization in these lines. 
Other elements have been identified in polarization spectra; 
notably Fe~{\sc ii} lines in SNe 1997dt (Leonard et al. 2005), 
and 2004S (Chornock \& Filippenko 2008) and 
Mg~{\sc ii} in SN 2004dt (Wang et al. 2006) and SN 2006X (Patat et al. 2009).  
For a review of polarimetric studies of SNe~Ia, see Wang \& Wheeler (2008).

SN~2011fe occurred in M101, and was discovered on 2011 August 24 by
the Palomar Transient Factory (PTF: Nugent et al. 2011a). 
The proximity of M101, $\sim$6.2 Mpc, and indications that the SN 
suffered minimal host galaxy extinction suggested
that the SN would become the brightest SN~Ia since SN~1972E. 
Studies of the
light curve suggest that the SN was discovered just 0.5 days after the
explosion, and the explosion time is constrained to very high
precision (Nugent et al. 2011b). Optical spectra revealed SN~2011fe
to be a normal SN~Ia, with detections of C~{\sc ii} $\lambda$6580 \AA\ and
$\lambda$7234 \AA\ in absorption (Cenko et al.\ 2011; Parrent et al. 2012; 
Pereira et al. 2013), 
and the blueshift of Si~{\sc ii} $\lambda$6355 \AA\ placed it in the 
low-velocity (LVG/LV) class (see Benetti et al. 2005, Wang et al. 2009, 
Foley et al. 2011 for LVG/HVG and LV/HV definitions). 
Studies of pre-explosion images of the site of SN~2011fe place the strictest
upper limits yet on the luminosity of any SN~Ia progenitor, arguing
against a single-degenerate progenitor containing a giant donor star
(Li et al.\ 2011). Observations of the SN in the 
optical (Nugent et al. 2011b; Bloom et al. 2012) and 
UV/X-ray wavelength ranges (Brown et al. 2012) constrained the 
exploding star to be smaller than main sequence stars, and further 
constrained the donor stars. The UV-optical colors demonstrated that 
the SN is of the NUV-blue subset of normal SNe~Ia (Milne et al. 2013). 
Patat et al. (2013) report a small change in a 
component of the Na~I~D absorption for SN~2011fe between +3 days and +19 days, 
a change that has been attributed to the presence of circumstellar material (CSM) 
(Patat et al. 2007; Simon et al. 2009). 
However, by investigating reasonable expectations of the change of this 
absorption line with the geometrical increase in the emitting 
region compared to the size scale of variations in ISM, Patet et al. (2013) found 
that the variation could be due to ISM and not CSM. 

SN~2011fe was the nearest Type~Ia explosion in several decades,
providing an unprecedented opportunity to obtain spectropolarimetry of
a normal LV SN~Ia with modest-aperture telescopes. We initiated a
campaign to obtain multi-epoch spectropolarimetry of SN~2011fe at
Steward Observatory, using the 1.5-m Kuiper, the 2.3-m Bok
and the 6.5-m MMT telescopes.  We describe the results of these observations below.

\begin{figure*}
\epsscale{1.0} 
\plotone{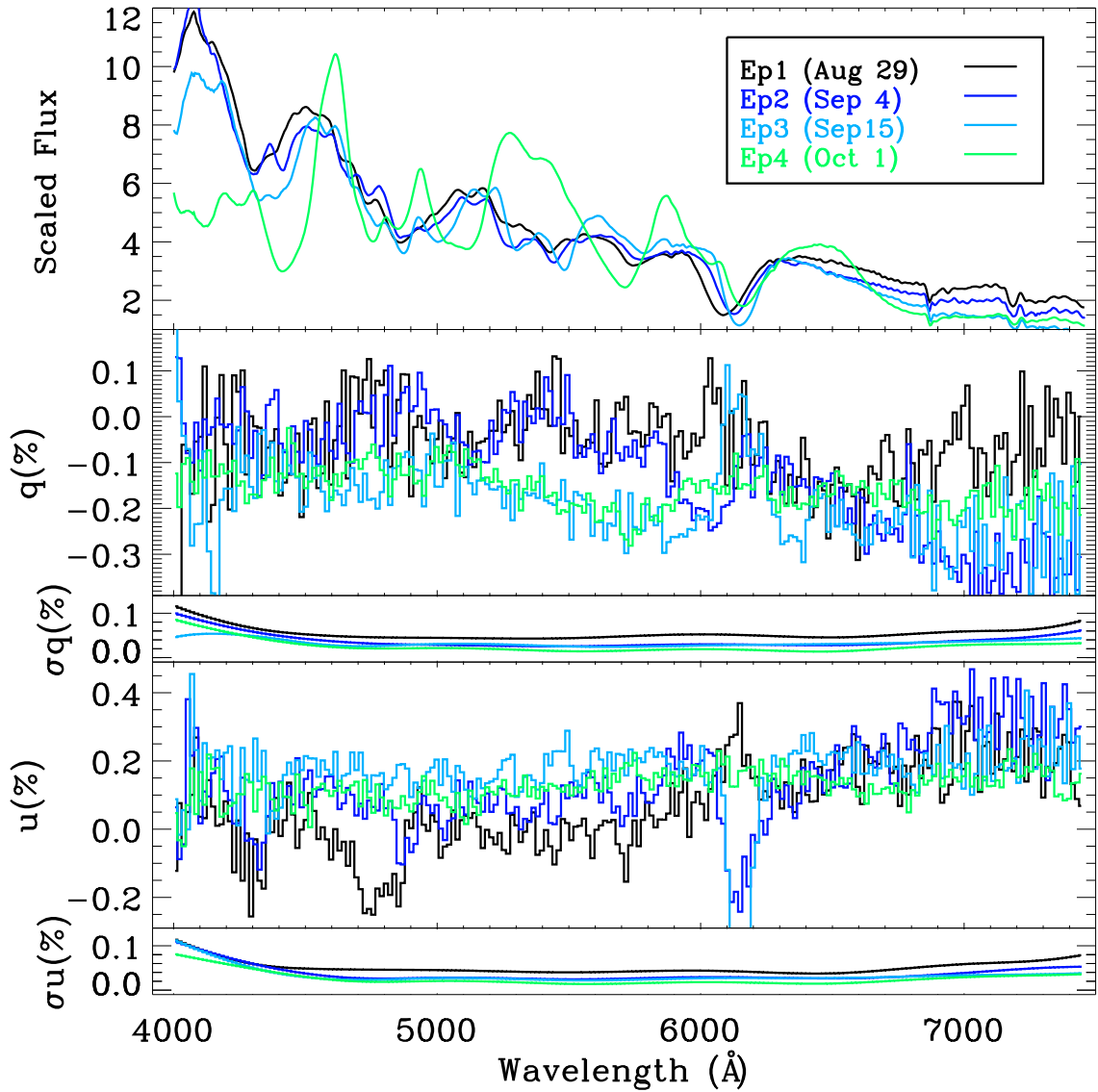}
\caption{The first four epochs (color coded) of flux, 
  and uncorrected Stokes parameters $q$ and $u$.  The
  flux values have been scaled to the first epoch flux at 5800 \AA \
  using the following flux ratios: 1.0, 0.228, 0.148, 0.215. The 
 $q$ and $u$ spectra have been binned to 16 \AA\ bins, with the corresponding 
errors shown below each spectrum.}
\label{fig:raw_flx_qu}
\end{figure*}

\begin{table}\begin{center}\begin{minipage}{3.50in}
%\begin{table*}\begin{center}
      \caption{SPOL Observation log for SN~2011\lowercase{fe}}
\scriptsize
\tighten
\begin{tabular}{@{}lcccccc}\hline\hline
%UT Date &Age &Age & UT & Telescope\tablenotemark{c} & Exp  &Epoch \\
%             &+exp\tablenotemark{a} &  $+$B$_{\rm
%               max}$\tablenotemark{b} & (start) & &  & \\
UT Date &Age &Age & UT & Telescope$^{c}$ & Exp  &Epoch \\
             &+exp$^{a}$ &  $+$B$_{\rm max}^{b}$ & (start) & &  & \\
 & (days) & (days) & & & (sec) & \\
\hline
2011-08-29 & 5.4  & -12.0 & 03:32:02  &  Bok & 3840  & 1      \\
2011-09-04 & 11.4 & -6.0 & 02:51:44  &  Kuiper & 4320  &2 \\
2011-09-15 & 22 &  5 & 02:53:41  &  Bok & 1504 & 3      \\
2011-09-16 & 23 &  6 & 02:48:08  &  Bok & 2784 & 3      \\   
2011-09-26 & 33 & 16 & 02:41:28  &  Bok & 1440 & 4      \\
2011-09-28 & 35 & 18 & 02:53:44  &  Bok & 800  & 4      \\ 
2011-09-29 & 36 & 19 & 02:20:22  &  Bok & 1920 & 4      \\
2011-09-30 & 37 & 20 & 02:26:08  &  Bok & 1920 & 4      \\ 
2011-10-01 & 38 & 21 & 02:24:33  &  Bok & 1280 & 4      \\ 
2011-10-06 & 43 & 26 & 02:09:04  &  Bok & 1280 & 4      \\
2011-11-28 & 96 & 79 & 12:38:26  &  Kuiper & 320 & 5 \\
2011-12-27 & 125 & 108 & 11:57:12 & Kuiper & 720 & 6 \\
2011-12-28 & 126 & 109 & 12:32:48 & Kuiper & 480 & 6 \\
2011-12-29 & 127 & 110 & 12:08:05 & Kuiper & 720 & 6 \\
2011-12-30 & 128 & 111 & 11:18:47 & Kuiper & 720 & 6 \\
2012-01-01 & 129 & 112 & 11:33:31 & Kuiper & 720 & 6 \\
2012-01-26 & 155 & 138 & 11:26:55 & Bok & 960 & 7 \\
2012-01-28 & 156 & 139 & 13:04:04 & Bok & 360 & 7 \\
2012-02-15 & 174 & 157 & 12:47:13 & Bok & 360 & 8 \\
2012-02-21 & 180 & 163 & 12:34:54 & Bok & 360 & 8 \\
2012-03-25 & 213 & 196 & 11:41:57 & Bok & 480 & 9 \\
2012-04-16 & 235 & 218 & 10:28:33 & MMT & 960 & 10    \\

\hline
\end{tabular}
%\tablenotetext{a}{Epoch relative to explosion, 23.7 Aug. 2011, as
%  estimated by averaging explosion dates of Brown et al. (2011) and
%  Nugent et al. (2011).}  
%\tablenotetext{b}{Epoch relative to the reported date of B$_{\rm
%    max}$, 10.1 $\pm$0.2 Sep. 2011, as reported by Matheson et al.
%  (2011).}  
%\tablenotetext{c}{$Bok$ denotes the 2.3-m Bok telescope on Kitt Peak, AZ. 
%$Kuiper$ denotes the 1.54-m Kuiper telescope at Mt. Bigelow, AZ. 
%$MMT$ denotes the 6.5-m MMT telescope at Mt. Hopkins, AZ. The same SPOL instrument 
%was used for all observations.}
\begin{tabular}{l}
$^{a}$ Epoch relative to explosion, 23.7 Aug. 2011, as
estimated by \\ 
averaging explosion dates of Brown et al. (2011) and
Nugent et al. \\
(2011). \\
$^{b}$ Epoch relative to the reported date of B$_{\rm max}$, 10.1 
$\pm$0.2 Sep. 2011, \\
as reported by Matheson et al. (2011). \\
$^{c}$ $Bok$ denotes the 2.3-m Bok telescope on Kitt Peak, AZ. $Kuiper$ \\
denotes the 1.54-m Kuiper telescope at Mt. Bigelow, AZ. $MMT$ \\
denotes the 6.5-m MMT telescope at Mt. Hopkins, AZ. The same \\
SPOL instrument was used for all observations.\\
\end{tabular}
\label{tab:obslog}
%\end{center}\end{table*}
\end{minipage}\end{center}\end{table}

\begin{figure*}
\epsscale{1.0}
\plotone{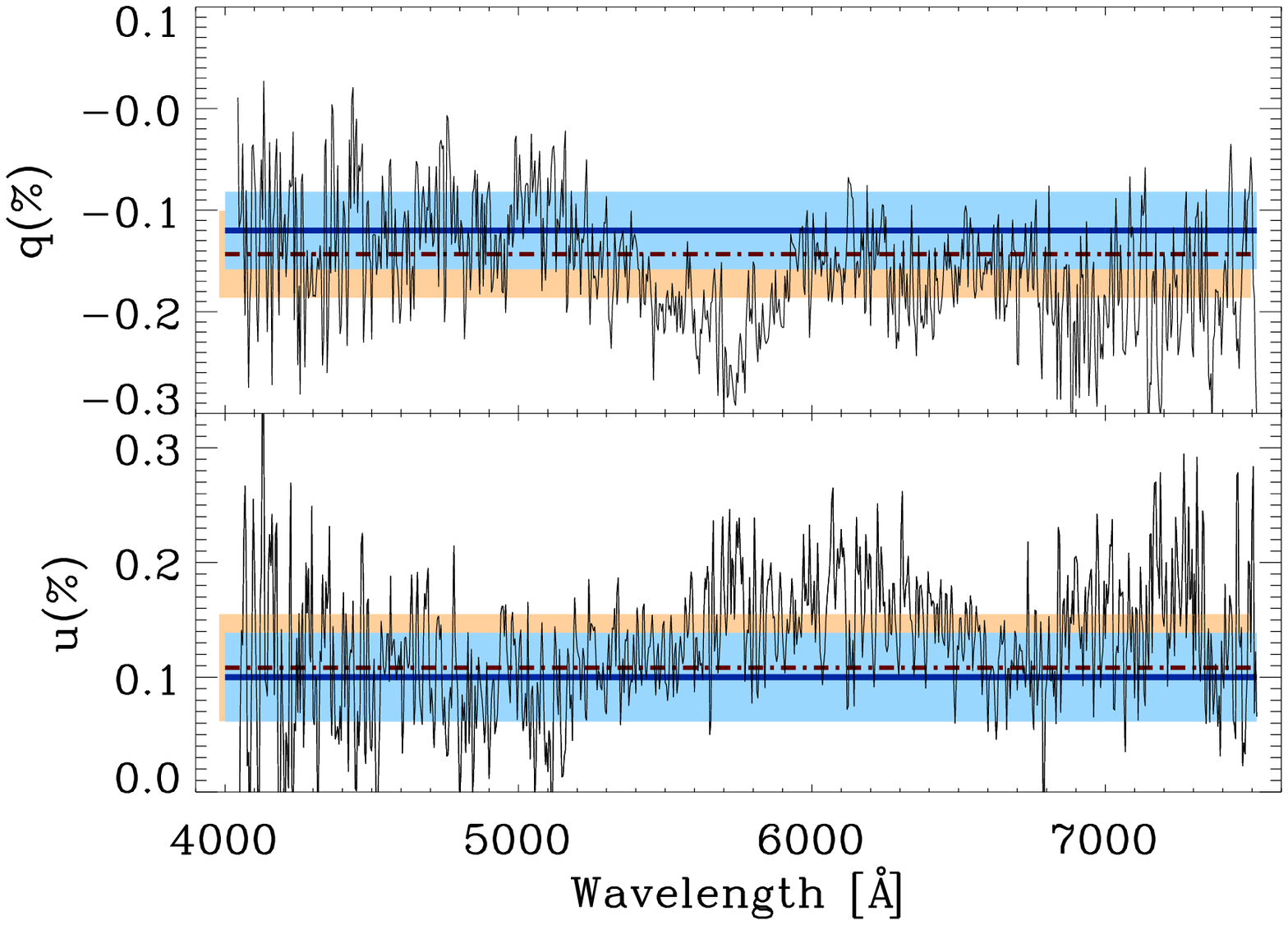}
\caption{$q$ \& $u$ spectra of SN~2011fe from Epoch 4 
compared to values chosen for $q$(ISP) and $u$(ISP).
The solid blue line shows the weighted average of the 
4600 - 5400 \AA wavelength range for Epochs 4-10. 
The dot-dashed red line shows the average from the wider 
4000-7000 \AA\ wavelength range for Epochs 4-10. The blue and 
red shaded regions show the standard deviations of the 
individual epoch averages compared to the overall 
average values.  }  
\label{fig:qisp_uisp}
\end{figure*}

\begin{figure*}
\epsscale{1.0}
\plotone{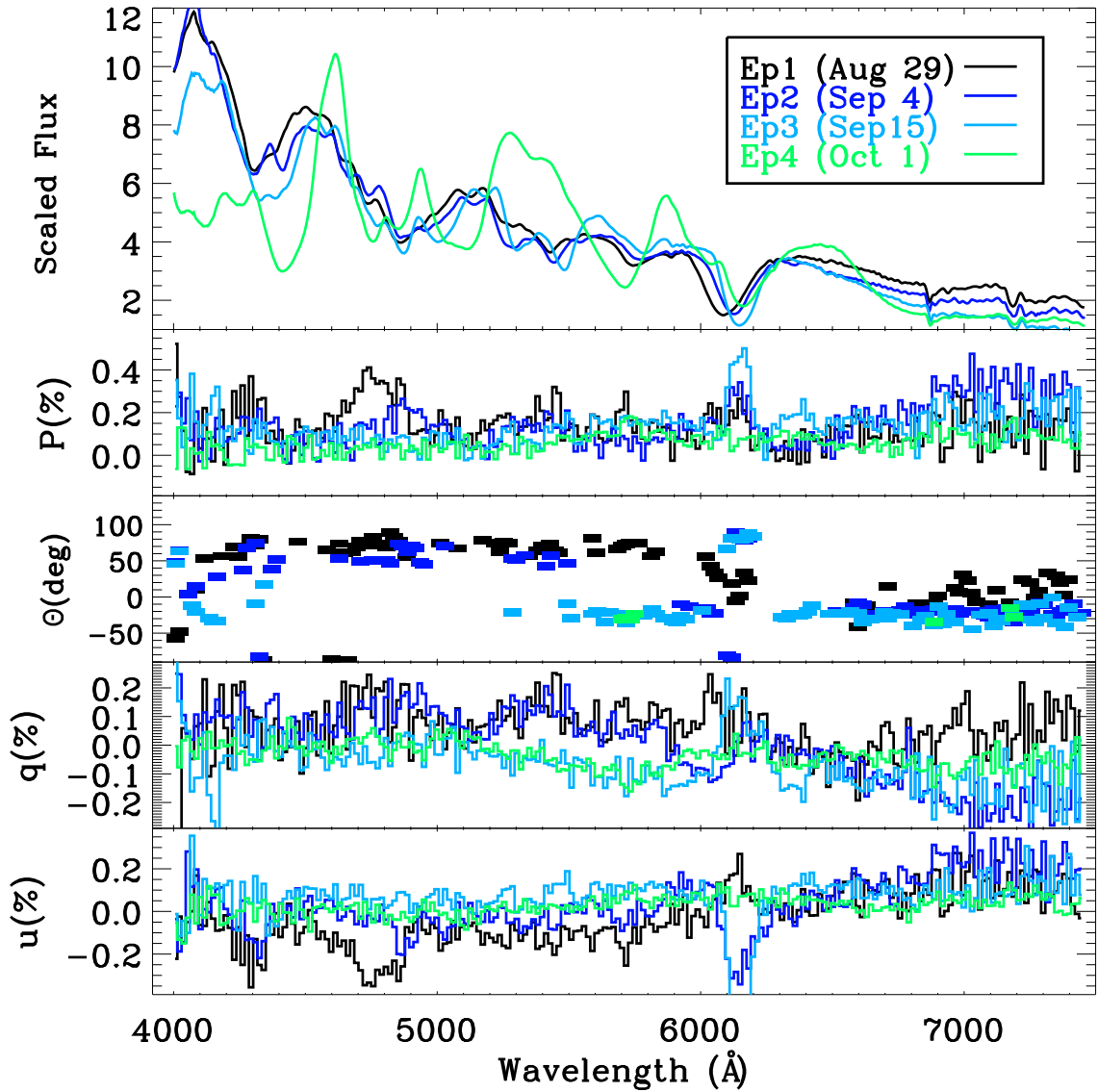}
\caption{The first four epochs (color coded) of flux, degree of
polarization, angle, and ISP-corrected Stokes parameters $q$ and $u$.  The
flux values have been scaled to the first epoch flux at 5800 \AA \
using the following flux ratios: 1.0, 0.228, 0.148, 0.215. P and $\theta\/$ 
were determined with the ISP-corrected $q$ and $u$. For presentation purposes, 
$\theta\/$ was only plotted when $P > 0.08\%$. 
The P,~$\theta\/$, $q$ and $u$ spectra have been binned into 16 \AA\ bins.} 
\label{fig:spec-seq}
\end{figure*}

\begin{figure*}
\epsscale{0.75} 
\plotone{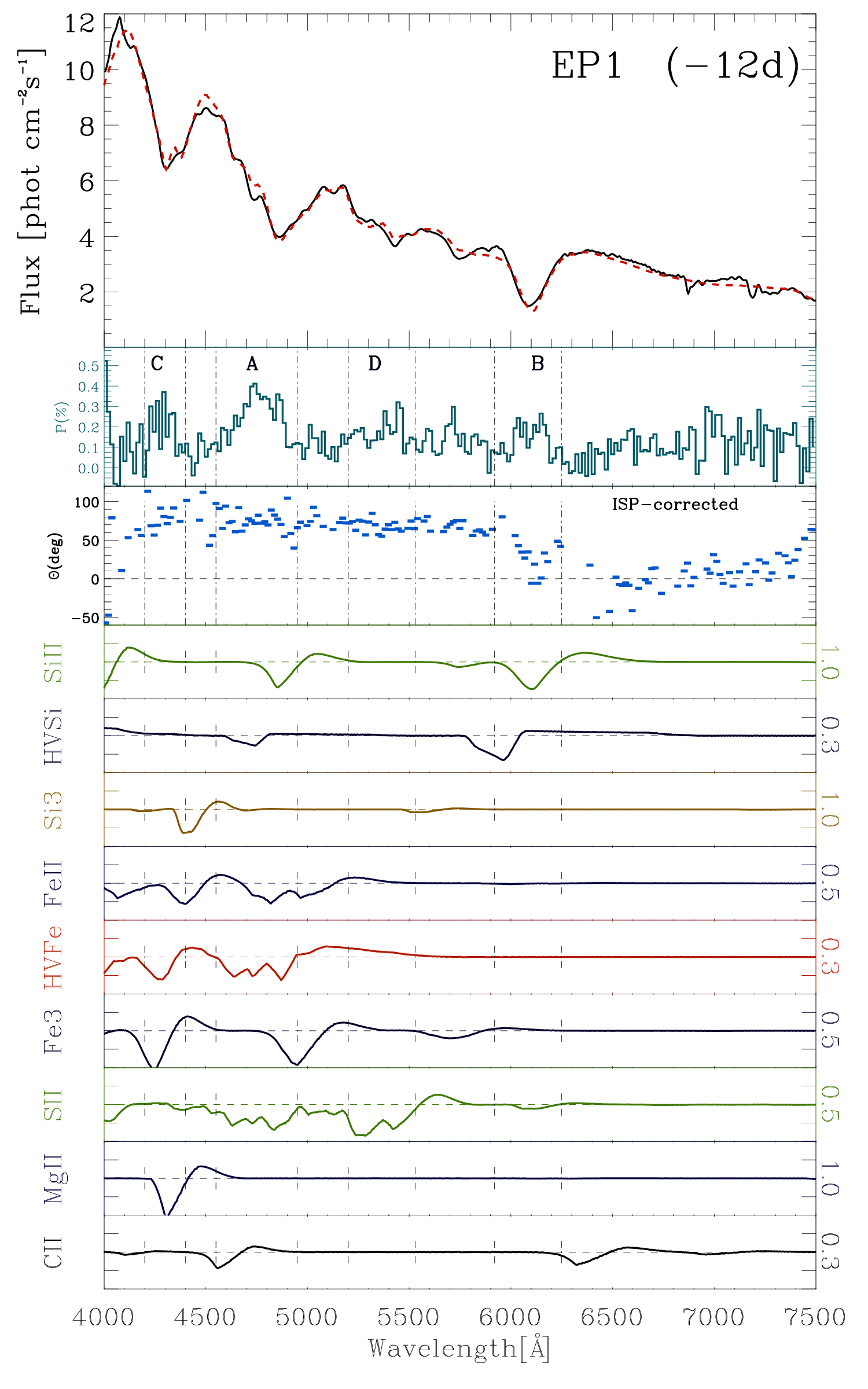}
\caption{Upper panel: Flux spectrum of Epoch 1 fit with $SYNAPPS$ algorithm. 
The flux spectrum is shown with a black solid line, the $SYNAPPS$  
fit is shown as a red dashed line. Fit parameters are listed in 
Table \ref{synparams}. Lower panels: Polarization and $\theta\/$ spectra compared 
against individual ions from $SYNAPPS$ fits for Epoch 1. The ion flux values 
are plotted on different scales, as shown on the right axis. The ISP-corrected 
P and $\theta\/$ spectra have been binned into 16 \AA\ bins. 
See text for explanation of $A$,$B$,$C$,$D$ labels.}
\label{fig:synapps1}
\end{figure*}

\section{OBSERVATIONS}

The CCD Imaging/Spectropolarimeter (SPOL; Schmidt et al. 1992a) mounted on
the Steward Observatory 2.3-m Bok telescope (Kitt Peak, AZ) the
1.54-m Kuiper telescope (Mt.\ Bigelow, AZ), and the 6.5-m MMT telescope 
(Mt. Hopkins, AZ) was used to obtain spectropolarimetry of SN~2011fe 
on 22 nights over 8 months. We have grouped those nights into 10 epochs in 
Table~1. Grouping was justified by a confirmed lack of inter-night 
variations in the $q$ or $u$ spectra during each epoch. 
%Observations
%covered 4000--7550 \AA\ at a resolution of $\sim$20~\AA\ (600 line
%mm$^{-1}$ grating in first order, using a
%5$\farcs$1$\times$51$\arcsec$ \ slit and a Hoya L38 blocking filter).
Observations at the Kuiper and Bok telescopes used the 600 line mm$^{-1}$  
grating in first order with a 5.1" x 51" slit and a Hoya L38 blocking filter 
providing a spectral resolution of $\sim$20\AA\ from 4000-7550\AA.  
Observations at the MMT used the 964 line mm$^{-1}$ grating in first order 
with a 1.9" x 19" slit and a Hoya L38 blocking filter providing a spectral 
resolution of $\sim$20\AA\ from 4100-7200\AA.
A rotatable semiachromatic half-wave plate was used to modulate
incident polarization and a Wollaston prism in the collimated beam
separated the orthogonally polarized spectra onto a thinned,
anti-reflection-coated 800$\times$1200 SITe CCD.  The efficiency of
the wave plate as a function of wavelength is measured by inserting a
fully-polarizing Nicol prism into the beam above the slit.  A series
of four separate exposures that sample 16 orientations of the wave
plate yields two independent, background-subtracted measures of each
of the normalized linear Stokes parameters, $q\/$ and $u\/$.  Each
night, several such sequences of observations of SN~2011fe were
obtained and combined, with the weighting of the individual
measurements based on photon statistics. 

We confirmed that the instrumental polarization of SPOL mounted on the
Bok, Kuiper and MMT telescopes is much less than 0.1\% through observations
of the unpolarized standard stars BD+28$^{\circ}$4211, HD~212311 and 
G191B2B (Schmidt et al. 1992b). The adopted correction
from the instrumental to the standard equatorial frame for the 
linear polarized position angle on the sky ($\theta\/$) 
for all epochs was determined from the average position angle offset
of Hiltner~960 and VI~Cyg~\#12. 
Additional observations of the polarization standard stars 
BD+59$^{\circ}$389 and BD+64$^{\circ}$106 were made during the third epoch 
and HD 245310 during the last epoch. Differences between the measured and expected 
polarization position angles were $<$ 0.5 degrees for all of the standard stars.

During Epoch 1, two field stars within $\sim$2\arcmin \ of
SN~2011fe (2MASS J14031367+5415431 and 2MASS J14025413+5416288) were
measured to check for significant Galactic interstellar polarization
(ISP) along the line-of-sight to the SN.  These stars yielded a
consistent estimate for Galactic ISP, with $P_{max} = 0.11 \pm 0.03$\%
at $\theta = 114^{\circ} \pm 7^{\circ}$ for 2MASS J14031367+5415431
and $P_{max} = 0.16 \pm 0.03$\% at $\theta = 109^{\circ} \pm
6^{\circ}$ for 2MASS J14025413+5416288, assuming that $\lambda_{max}$,
the wavelength where the ISP is at a maximum
($P_{max}$) is 5550 \AA (Serkowski, Mathewson \& Ford 1975). 
The results for the field stars were averaged
and $P_{max} = 0.13$\% at $\theta = 112^{\circ}$ was adopted as the
Galactic ISP in the sightline to SN~2011fe.  This low value for the
Galactic ISP is consistent with the high Galactic latitude of M101 and
the very low estimated amount of extinction, E(B-V)$\leq$0.1 mag, 
for the supernova (Foley \& Kirshner 2013).  
%The polarization spectra of SN~2011fe have been corrected for this level
%of Galactic ISP assuming that it is fit well by a Serkowski curve 
%\citep{wilking80,serkowski}.  
We reduced the data using custom, but mature IRAF routines. We began by 
bias-subtracting and flat-fielding each image, and used observations 
of He, Ne, and Ar lamps at the beginning of each run for wavelength 
calibration purposes. We then extracted Stokes paramters $q$ and $u$. 
We de-biased the positive definite nature of the polarization calculation 
using the prescription 
$P = \pm \sqrt{\mid Q^2 + U^2 - \frac{1}{2}(\sigma^2_Q + \sigma^2_U)\mid}$ 
(Wardle \& Kronberg 1974) where 
the sign is determined according to the sign of the modulus. Finally, 
to further increase the signal-to-noise ratio, we binned the Stokes parameters, 
to 16 \AA ~wide intervals. 

Since our standard instrumental set-up for the SPOL instrument 
does not reach to long enough wavelengths, this work does not include 
any observations of the Ca IR lines, which are strongly polarized in 
some SNe~Ia. The standard set-up is chosen to avoid the significant fringing 
present at wavelengths longer than 7000 \AA~.

\begin{figure*}
\epsscale{0.75}
\plotone{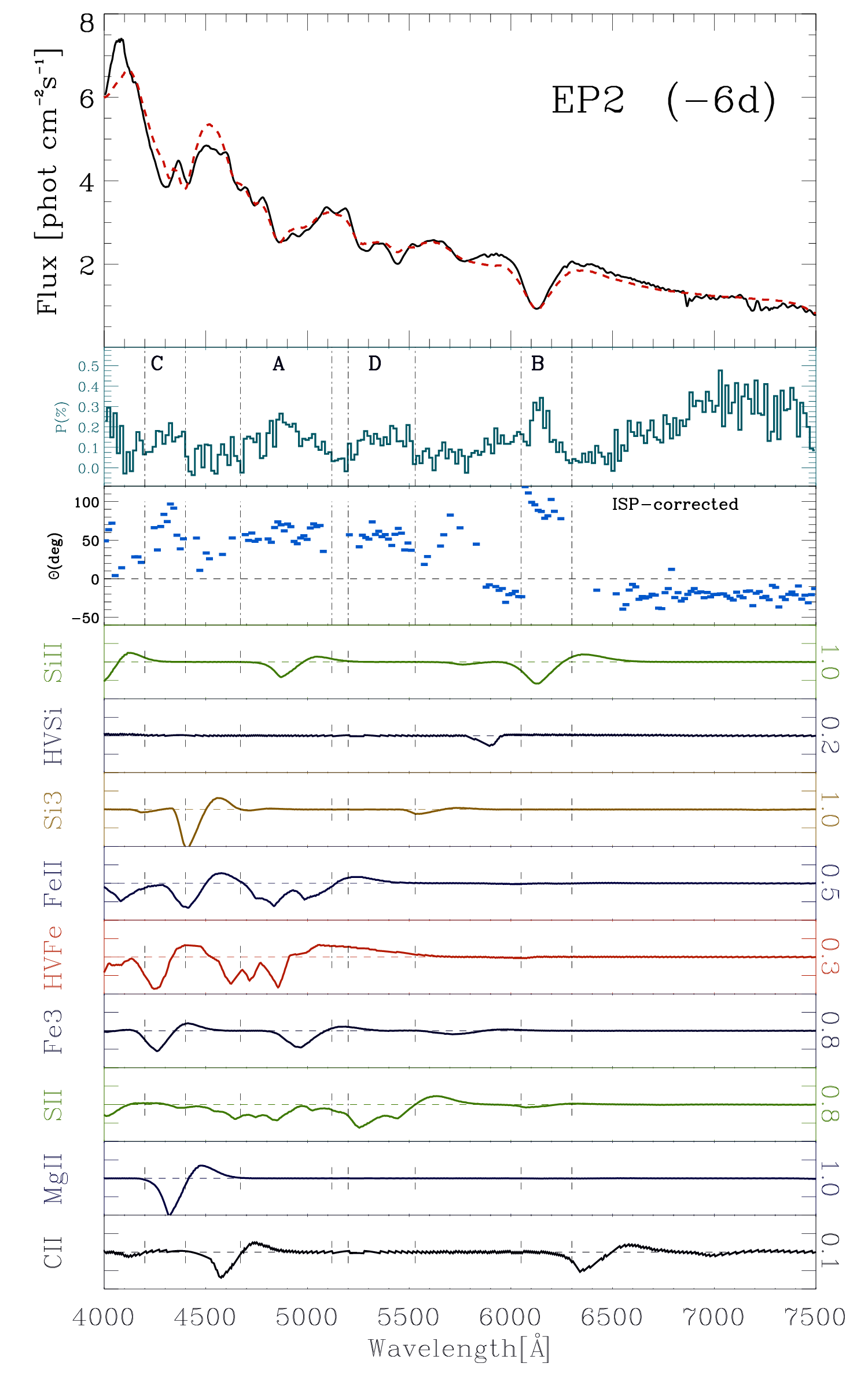}
\caption{Upper panel: Flux spectrum of Epoch 2 fit with $SYNAPPS$ algorithm.
The flux spectrum is shown with a black solid line, the $SYNAPPS$  
fit is shown as a red dashed line. Fit parameters are listed in
Table \ref{synparams}. Lower panels: Polarization and $\theta\/$ spectra 
compared
against individual ions from $SYNAPPS$ fits for Epoch 2. The ion flux values 
are plotted to different scales, as shown on the right axis.
See text for explanation of $A$,$B$,$C$,$D$ labels. }
\label{fig:synapps2}
\end{figure*}

\begin{figure*}
\epsscale{0.75}
\plotone{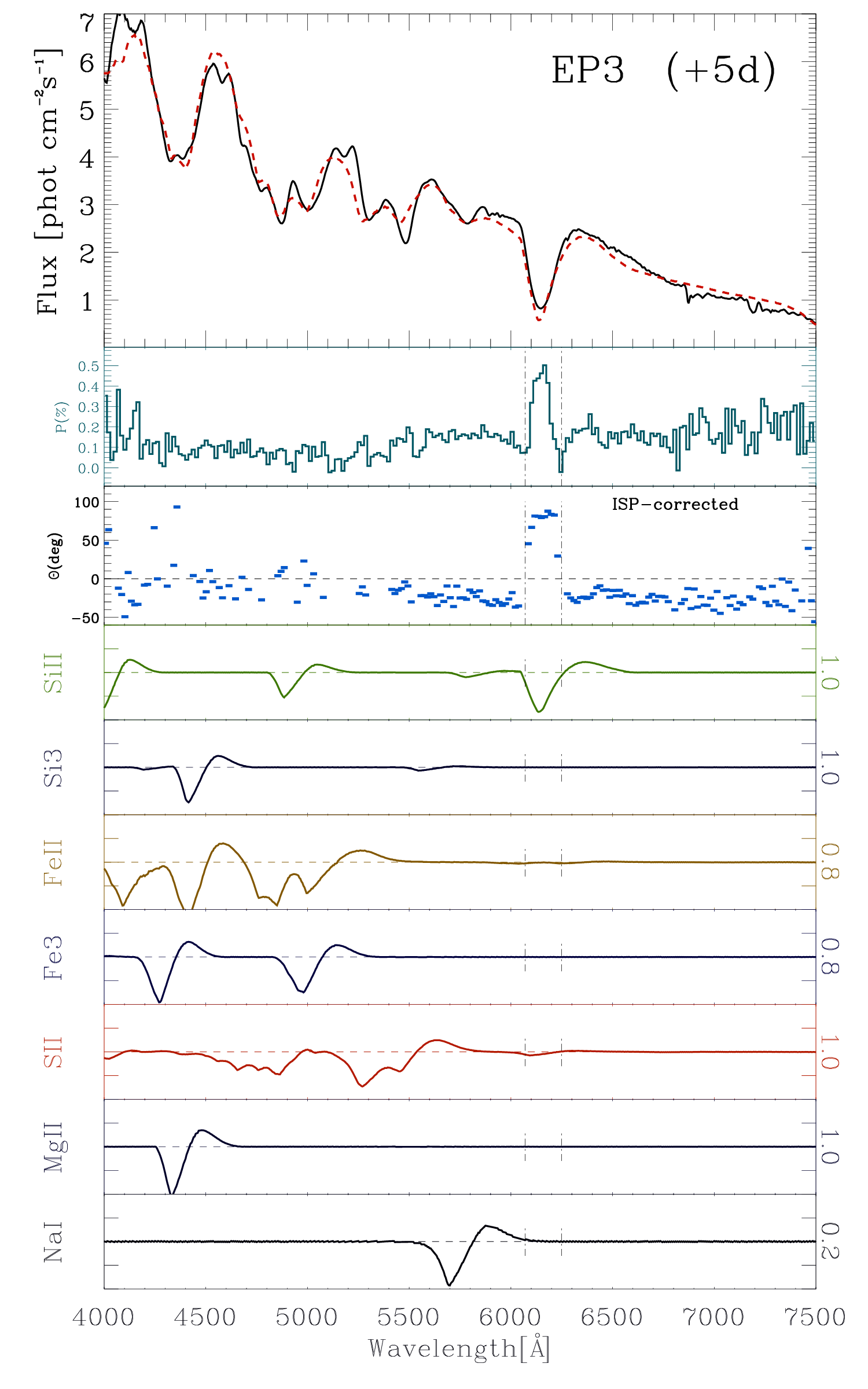}
\caption{Upper panel: Flux spectrum of Epoch 3 fit with $SYNAPPS$ algorithm.
The flux spectrum is shown with a black solid line, the $SYNAPPS$  
fit is shown as a red dashed line. Fit parameters are listed in
Table \ref{synparams}. Lower panels: Polarization and $\theta\/$ spectra compared
against individual ions from  $SYNAPPS$ fits for Epoch 3. The ion flux values 
are plotted to different scales, as shown on the right axis.
See text for explanation of $A$,$B$,$C$,$D$ labels.}
\label{fig:synapps3}
\end{figure*}

\begin{table}\begin{center}\begin{minipage}{3.1in}
      \caption{Synapps fit parameters.}
\scriptsize
\tighten
\begin{tabular}{lccccc}\hline\hline
Ion & $\tau$ & $v_{min}$ &       $v_{max}$ &     $v_{e}$     & T \\
%   &         & [10,000 km/s] & [10,000 km/s] & [10,000 km/s] & [1000 K]\\
   &         & [10$^{4}$ km/s] & [10$^{4}$ km/s] & [10$^{4}$ km/s] & [10$^{3}$ K]\\
\hline
\multicolumn{6}{c}{Epoch 1} \\
\hline
SiII & 22.4 & 12.5 & 26.0 & 2.0 & 10 \\ 
SII  & 3.31 & 12.5 & 15.1 & 2.0 & 10 \\
FeII & 1.38 & 12.5 & 23.1 & 2.0 & 10 \\
CII  & 0.05 & 12.5 & 30.0 & 2.0 & 10 \\
MgII & 3.72 & 12.5 & 17.9 & 2.0 & 10 \\
SiIII & 3.09 & 12.5 & 15.0 & 2.0 & 10 \\
FeIII & 2.88 & 12.5 & 25.5 & 2.0 & 10 \\
HV SiII & 0.71 & 19.5 & 30 & 7.13 & 10 \\  
HV FeII & 0.47 & 18.6 & 30 & 6.01 & 10 \\
\hline
\multicolumn{6}{c}{Epoch 2} \\
\hline
SiII & 5.62  & 11.4 & 26.3 & 2 & 10 \\
SII  & 1.38  & 11.4 & 30 & 2   & 10 \\
FeII & 0.83  & 11.4 & 21 & 2   & 10 \\
CII  & 0.01  & 11.4 & 23.6 & 2 & 10 \\
MgII & 1.66  & 11.4 & 23.1 & 2 & 10 \\
SiIII & 2.0  & 11.4 & 15   & 2 & 15 \\
FeIII & 1.12 & 11.4 & 18.8 & 2 & 15 \\
HV SiII & 3.16 & 23.3 & 30 & 3.22 & 10 \\
HV FeII & 0.63 & 19.6 & 30.0 & 6 &  10 \\
\hline
\multicolumn{6}{c}{Epoch 3} \\
\hline
SiII & 12.6 & 10.5 & 15 & 2 & 7 \\
SII &  1.45 & 10.5 & 28 & 2 & 7 \\
FeII & 2    & 10.5 & 20 & 2 & 7 \\
CII & ---  & --- & --- & --- & --- \\
MgII & 1.32 & 10.5 & 16.2 & 2 & 7 \\
SiIII & 0.83 & 10.5 & 15 & 2 & 10 \\
FeIII & 0.74 & 10.5 & 15 & 2 & 10 \\
NaI & 0.47  & 10.5 & 26.9 & 2 & 7 \\
\hline
\end{tabular}
\begin{tabular}{ll}
\end{tabular}
\end{minipage}
\end{center}
\label{synparams}
\end{table}

\section{POLARIZATION OF SN~2011\lowercase{fe}: Peak Epochs 1-4}

Our sequence of the first 4 epochs of spectra are shown in the top panel of
Figure~\ref{fig:raw_flx_qu}, displaying the emergence of absorption
features typical of SNe~Ia. The lower panels show the $q$ and $u$ 
polarization spectra, uncorrected for either Galactic polarization or 
host galaxy ISP. Epochs 1-3 show clear 
signatures of line polarization in both $q$ and $u$ spectra, while 
Epoch 4 appears to show no line polarization features. 
By Epoch 4 the spectrum is approaching the nebular phase, a time when 
the optical depth to electron scattering is low and thus the intrinsic 
SN polarization is expected to be low. Polarization observed in the 
nebular epoch is often assumed to be due to ISP. However, 
Figure \ref{fig:qisp_uisp} shows that there remains some level of 
intrinsic line polarization at Epoch 4, leading us to 
employ a different method in this work, 
computing the inverse variance-weighted average of $q$ and $u$ for the 
line-free continuum region spanning  
4600 -- 5400 \AA\, for Epochs 4-10.
%comparing three overlapping 
%wavelength regions in the data for npochs 4-10, computing the inverse 
%variance weighted average for all epochs. The regions are 
%4500\AA -- 5500\AA,  4600\AA -- 5400\AA and 5000\AA - 6000\AA.
%, and the results are shown in Table \ref{tab_isp}. 
We find $q$(ISP)=-0.12$\pm$0.06\% and $u$(ISP)=0.10$\pm$0.07\%. 
The error bars were determined by calculating the standard deviations of the 
individual epoch averages compared to  $q$(ISP) and $u$(ISP) 
These values are shown as blue line and the uncertainties as blue shading in 
Figure \ref{fig:qisp_uisp}. For comparison, also shown in Figure 
\ref{fig:qisp_uisp} are the same values calculated across the 
entire 4000 - 7000 \AA\ wavelength range (red lines and pink shading); 
the values are similar within the error bars.  
These error bars will be used in determining 
the uncertainties for the P$_{SiII}$ evolution (Sec. 4).  

By using constants, we do not account for any possible wavelength dependence 
of the ISP, which we consider a minor effect because of the very small 
polarization levels. A small wavelength dependence would lead to 
our single ISP value being biased toward the value at 5000 \AA. 
Figure \ref{fig:spec-seq} shows the flux, polarization, position angle, 
$q$ and $u$ spectra after subtracting our determination of the total 
ISP, which includes both a Galactic and host-galaxy contribution. 
We did not attempt to separate Galactic and host ISP in this work, but 
we note that our estimates of the Galactic ISP, shown in Section 2 are close 
to the total ISP we derive above. 
For presentation purposes, $\theta$ is only plotted when $P > 0.08\%$.

\subsection{Spectral Fitting with SYNAPPS}

In order to study the nature of the line polarization, we employed the 
$SYNAPPS$ algorithm (Thomas et al. 2011, which was derived from the $SYNOW$ 
algorithm, Branch et al. 2005) to fit the three spectra that showed 
line polarization features. The input parameters of the $SYNAPPS$ fits 
are shown in Table \ref{synparams}. We will discuss the $SYNAPPS$ fitting 
before attempting to match line features with polarization features. 

The upper panels of Figures \ref{fig:synapps1} - \ref{fig:synapps3}
show the $SYNAPPS$ fit to the Epoch 1-3 flux spectra, providing acceptable
fits to these spectra, particularly in the region of line polarization
features. The middle panels show the polarization (P)
and position angle ($\theta\/$), and will be discussed in the 
next section. The lower panels of Figures \ref{fig:synapps1} - \ref{fig:synapps3} 
show the contributions of individual ions to the composite spectra, 
with the flux scaling factor listed along the right axes to provide 
a sense of line strength. Different scaling factors were used to permit a 
better sense of the wavelength range for weaker features, with the scaling 
factor defined as the range compared to that of the Si~{\sc ii} range.  

The Epoch 1 spectrum has strong Si~{\sc ii} (with both a photospheric and weaker 
high-velocity component), Si~{\sc iii} and Mg~{\sc ii} absorption features. 
Iron and sulfur lines provide a broad wavelength range of weaker absorption. 
Also detected is C~{\sc ii}, 
which is present in about a third of SNe~Ia and is interpreted to be a 
signature of unburned carbon (Parrent et al. 2011; Thomas et al. 2011;
Folatelli et al. 2012; Silverman et al. 2012). The Epoch 2 fitting is similar to 
Epoch~1, but with the HV Si~{\sc ii} feature much weaker. Epoch~3 presents fewer lines, 
as the HV components of Si~{\sc ii} and Fe~{\sc ii} are no longer required, nor is  
C~{\sc ii} required, but now Na~{\sc i} is required. Across the epochs, the blueshift of the
photospheric Si~{\sc ii} feature decreases gradually with epoch, consistent with
a LV determination (Parrent et al. 2012). The multi-feature Fe~{\sc ii}, Fe~{\sc iii} 
and S~{\sc ii} complexes persist across the epochs. These $SYNAPPS$ fits can 
be compared to similar fits presented by Parrent et al. (2012) based upon optical 
spectra obtained as part of the Palomar Transient Factory program.

\subsection{Line Polarization} 

The middle panels of Figures \ref{fig:synapps1} -\ref{fig:synapps3}  
show the polarization (P) and position angle ($\theta\/$) for Epochs 1-3, 
only plotting $\theta$ values for polarizations above 0.08\%. 
There are 4 features present over the course of the 3 epochs, 
which we label Features $A$,$B$,$C$,$D$. 
Feature $A$ is a strongly polarized feature in the 4550-4950 \AA\ wavelength range during 
Epoch~1, but it fades so that by Epoch~3 the feature has disappeared. The 
position angle for Feature $A$ is consistent with the emission shortward of 6000\AA. 
Feature $B$ is in the 5900-6250 \AA\ wavelength range, a feature normally associated with the 
familiar Si~{\sc ii} $\lambda$6355 \AA\ absorption line. 
Feature $B$ 
strengthens so that by Epoch~3, it is the sole, dominant line feature. By Epoch~4, it 
has disappeared (Fig. 1). During Epoch~1, the polarization angle for Feature $B$ changes from 
$\sim$70 degrees to $\sim$0 degrees on either side of this feature. 
Feature $C$ is narrower, in the 4200-4400 \AA\ wavelength range, and evolves as if it 
is a weaker companion of Feature $A$. Feature $D$ is a broad, but initially poorly-defined feature 
in the 5000-5500 \AA\ wavelength range. The feature is most apparent during Epoch~2 and has 
disappeared by Epoch~3. 
Epoch~3 is notable both for the singular dominance of Feature $B$, but also because the 
polarization angle change between the blue and red halves of the spectrum 
that was seen in Epochs 1 and 2 has disappeared. 
We re-iterate that Epoch 4 was not fitted with $SYNAPPS$
as there are no line polarization features (Fig. \ref{fig:spec-seq}).

To understand the nature of these features, we compare them with the evolution of the individual 
spectral features, as explored with $SYNAPPS$.  The $SYNAPPS$ fitting supports that Feature $B$ is 
a Si~{\sc ii} feature. We accept that association and label Feature $B$ as $P_{SiII}$ throughout 
the remainder of this work. It is immediately clear that there are a few candidates to explain the 
polarization in Features $A$, $C$, and $D$.

Feature $A$ is located in a rapidly evolving portion of the spectrum where the absorption lines 
of S~{\sc ii}, Fe~{\sc ii}, and Si~{\sc ii} overlap.  HV Si~{\sc ii}  and photospheric Si~{\sc ii} 
are present over the indicated wavelength range in Epoch~1, but by Epoch~2 the absorption from 
HV Si~{\sc ii} is no longer evident. Perhaps this is the reason why the polarization is not as 
well defined at the bluer wavelengths in Epoch~2 as the opacity, and therefore the covering factor, 
of HV Si~{\sc ii} decreases. S~{\sc ii} and HV Fe~{\sc ii} show multiple absorption features 
over the wavelength range of Feature $A$ and could therefore be contenders. Photospheric Fe~{\sc ii} 
also has some absorption over this region, but the emission feature at shorter wavelengths must be 
included to explain the entire polarization feature in at least Epoch~1.

By examining the individual flux spectra of Epoch~1, we see that Mg~{\sc ii} and HV Fe~{\sc ii} are 
the ions with absorption features most contained within the defined wavelength interval of Feature $C$. 
Photospheric Fe~{\sc ii} and Fe~{\sc iii}, however, also have absorption features that at least 
partially fall across the wavelength range.  S~{\sc ii} is the only ion with absorption features 
associated with Feature $D$.  We note that photospheric and HV Fe~{\sc ii} and Fe~{\sc iii} have 
emission components across this range.

The elaborate blend of several elements in the flux spectrum of SN~2011fe does not allow us to 
determine with certainty if one particular element is the root of the ejecta asymmetries, however, 
we explore three possibilities, which we label as \emph{Interpretations 1-3}.  In \emph{Interpretation 1}, 
we assign Feature A to Si~{\sc ii} absorption from a complex of Si~{\sc ii} absorption lines 
(rest frame $\lambda5041$ \AA, $\lambda5056$ \AA, $\lambda5113$ \AA) seen in \emph{SYNAPPS} fits.  
If photospheric Si~{\sc ii} were the cause of line polarization at 4800 \AA, then each absorption feature 
of this element at optical wavelengths can be connected to line polarization features in Epochs 1 and 2.  
This includes the small feature at 5700 \AA ~that has been associated with Si~{\sc ii} 
$\lambda5958$ and $\lambda5979$ in SN~2004dt (Wang et al. 2006). However, to explain the broad polarization 
in Epoch~1 at 4800 \AA, we must include HV Si~{\sc ii} in Feature $A$.  Unfortunately the other absorption 
of this HV ion at 5900 \AA ~is not associated with line polarization which is a point of conflict for this 
interpretation.  Similarly, if Si~{\sc ii} explains both Features $A$ and $B$, then it is hard to explain 
the changing difference in angle between these two components.  Si~{\sc ii} is also not a viable candidate 
to explain Feature $C$ or $D$.  For \emph{Interpretation 1}, we speculate that Feature $A$ traces the 
same asymmetric layers in the SN atmosphere that are seen in Feature $B$ in Epoch 2 and 3, but at an 
earlier time.  This may be an important clue to the changes revealed by the receeding SN photosphere, 
which may be able to provide constraints on radiative transfer models of the observed polarization behavior.

\emph{Interpretation 2} assigns Feature $A$ to Fe~{\sc ii} and Fe~{\sc iii}.  
We note that the shifting absorption features of the iron complexes could explain why Feature $A$ is not as 
well defined in Epoch~2 as compared to Epoch~1.  Feature $C$ is then also explained by 
Fe~{\sc ii} and Fe~{\sc iii}, with the relative strengths 
between Features $A$ and $C$ matching the absorption rather well. HV Fe~{\sc ii} is also a potential contributor,  
although the second absorption from HV Fe~{\sc ii} 
of the $SYNAPPS$ fits shift to the blue and therefore cause part of the feature to move beyond the peak 
of the polarization feature.  A similar situation occurs for photospheric Fe~{\sc ii} and Fe~{\sc iii}, 
but the effect is to shift the absorption profiles into better sync with the polarization feature.  In general, 
the iron complexes absorb over a broad wavelength range and therefore make it difficult to match 
line polarization features with individual absorption features.

\emph{Interpretation 3} assigns Feature $A$ to S~{\sc ii}. There is strong absorption of 
S~{\sc ii} where the line polarization feature appears, but if S~{\sc ii} is also to explain Feature $D$, 
there is the issue that the polarization strength ratio does not 
match the ratio of the strength of the two broad absorption components.

These interpretations we outline above have been mentioned in previously published studies of 
spectropolarimetry of other SNe Ia (see Section 4).  There is not enough information to confidently 
distinguish between the three interpretations for SN~2011fe, but we are hopeful that this dataset can be 
studied in combination with other normal SNe Ia that we have observed to determine if any of the 
three possibilities are favored in general.

\begin{figure*}
\epsscale{0.60} 
%\plotone{q_u_6355.ps}
%\plotone{q_u_5051.ps}
\plotone{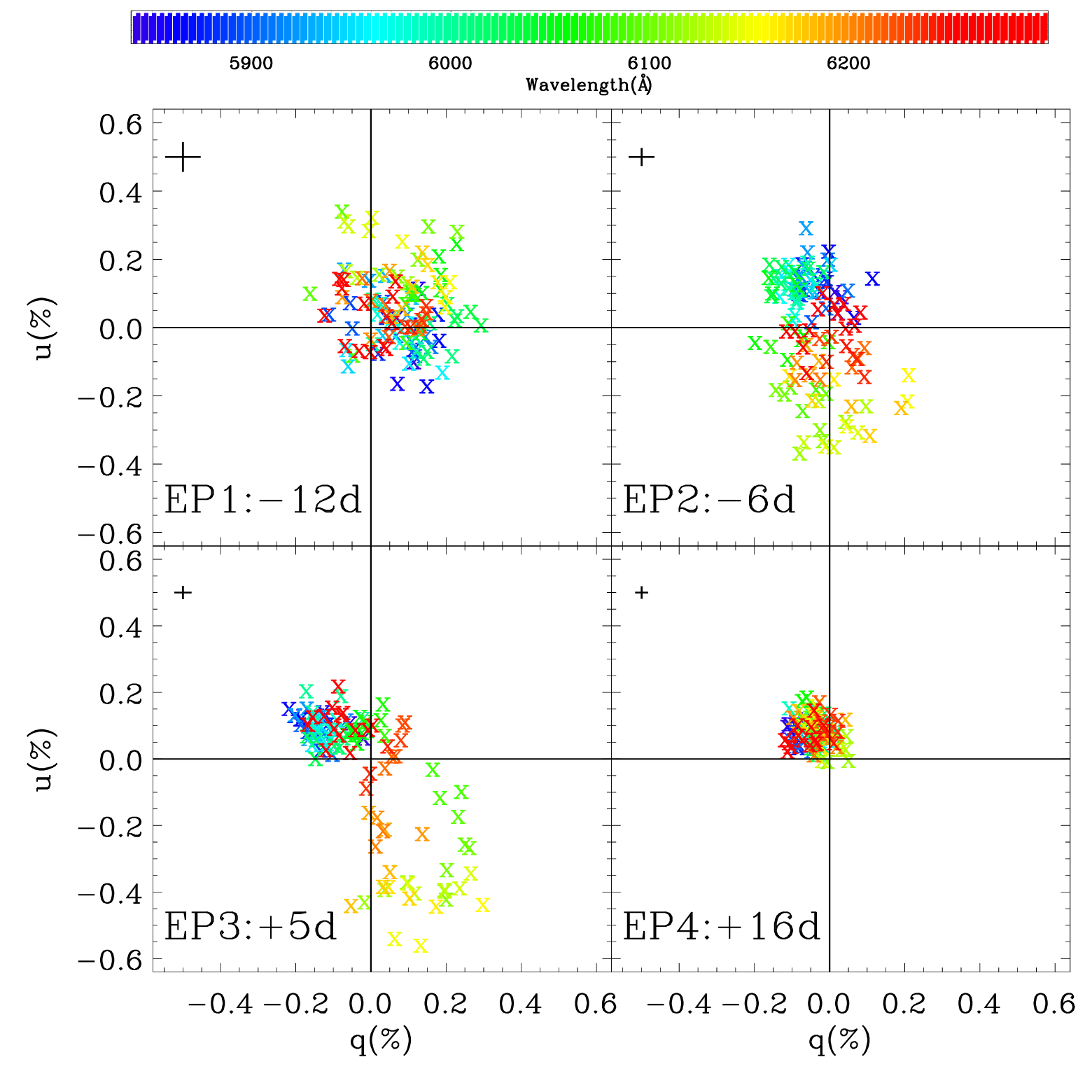}
\plotone{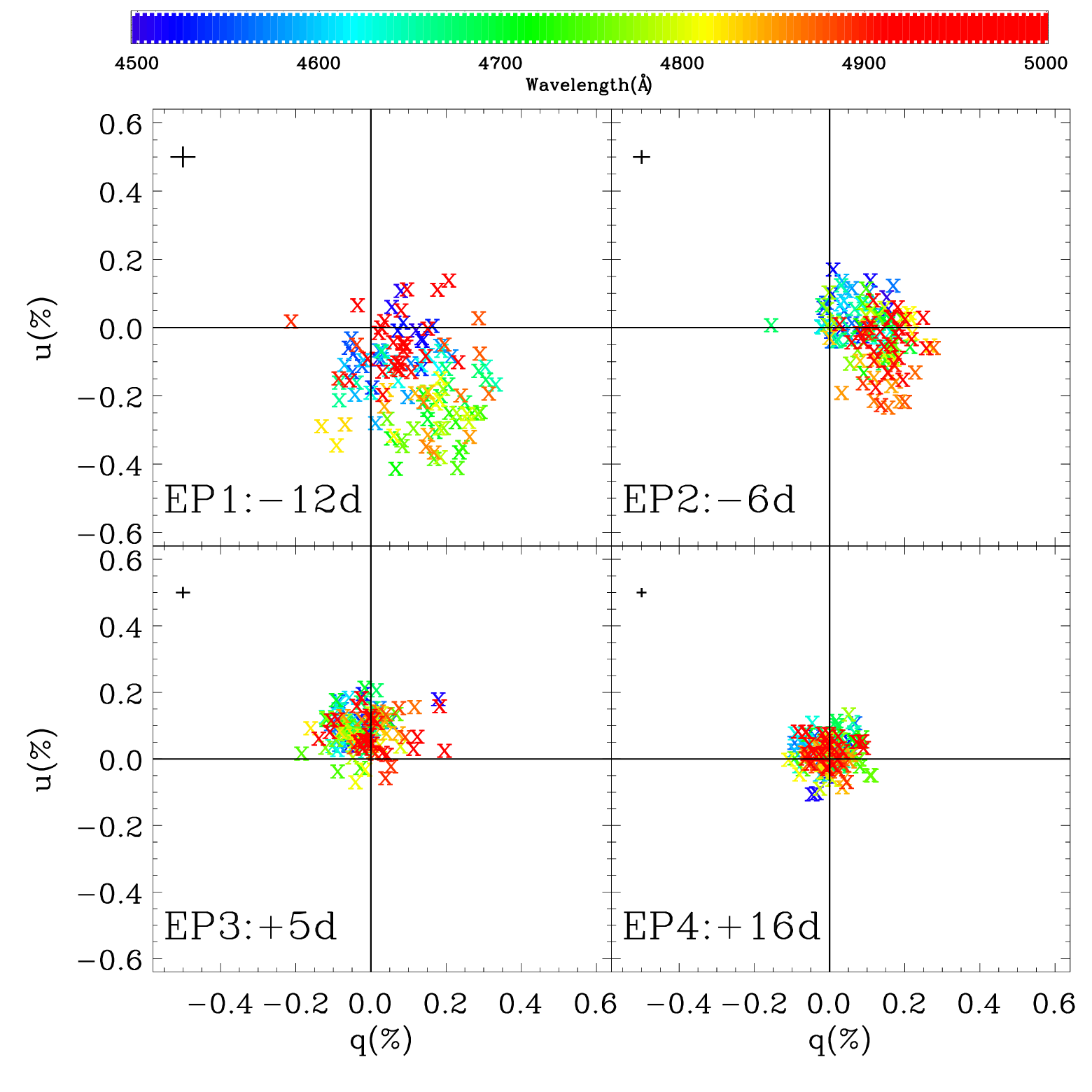}
\caption{Polarization features $B$ and $A$ plotted in the $q$/$u$ plane. 
The upper panel shows the $\lambda$6355 \AA\ line (feature $B$) for all four epochs, 
plotted from 5840 - 6300 \AA. The lower panel shows the feature $A$ for all 
four epochs, plotted from 4500 - 5000 \AA. 
The $\lambda$6355 \AA\ line exhibits clear loops during the second and third 
epochs, while there is a suggestion of multiple loops during Epoch 1 for 
feature $A$. The average error bar across each line is shown in the upper left of each 
panel.}
\label{fig:qu-loops}
\end{figure*}

Regardless of the interpretion of the source of the features, the line polarization 
can be further studied by plotting the evolution of Features $A$ and $B$ in $q-u$ space 
(Figure \ref{fig:qu-loops}). 
In Epochs 2 \& 3, the $\lambda$6355 \AA\ line is seen to exhibit a 
loop in the polarization (top panel). Interestingly, the loop appears 
counterclockwise when moving blue to red during Epoch~2, but the same
loop rotates clockwise in Epoch 3. 
As loops are suggested to be present when there is a deviation from 
axial symmetry, the Si~{\sc ii} distribution appears to deviate from 
spherical symmetry, but without exhibiting dominant axial symmetry. 
The lower panel shows the same presentation for Feature~$A$, in this case 
exhibiting a hint of multiple loops during Epoch~1 in the same quadrant as 
Feature~$A$. Since in all of the interpretations Feature~$A$ is actually a 
complex of lines, it is not clear that a clean loop should be expected.  

\begin{figure*}
\epsscale{0.70} 
\plotone{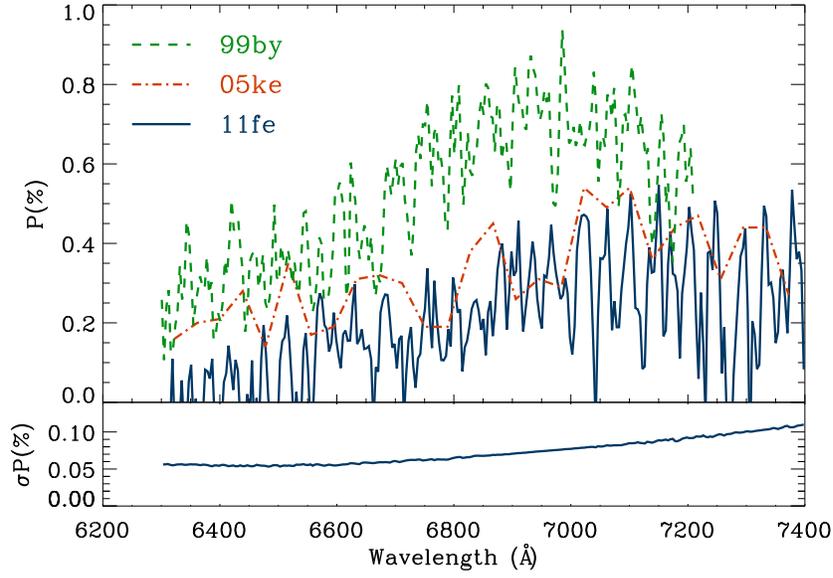}
\caption{Red continuum polarization of SN~2011fe (-6d) compared to 
SN~1999by (0d) and SN~2005ke (-8/-7d). SN~2011fe continuum polarization 
(solid blue line) roughly matches SN~2005ke 
(dot-dashed red line; Patat et al. 2012), 
and shows a similar increase with wavelength as SN~1999by 
(dashed green line; Howell et al. 2001). Errors of SN~2011fe data shown in 
lower panel.}
\label{vs_99by_05ke}
\end{figure*} 

\begin{figure*}
\epsscale{0.70} 
\plotone{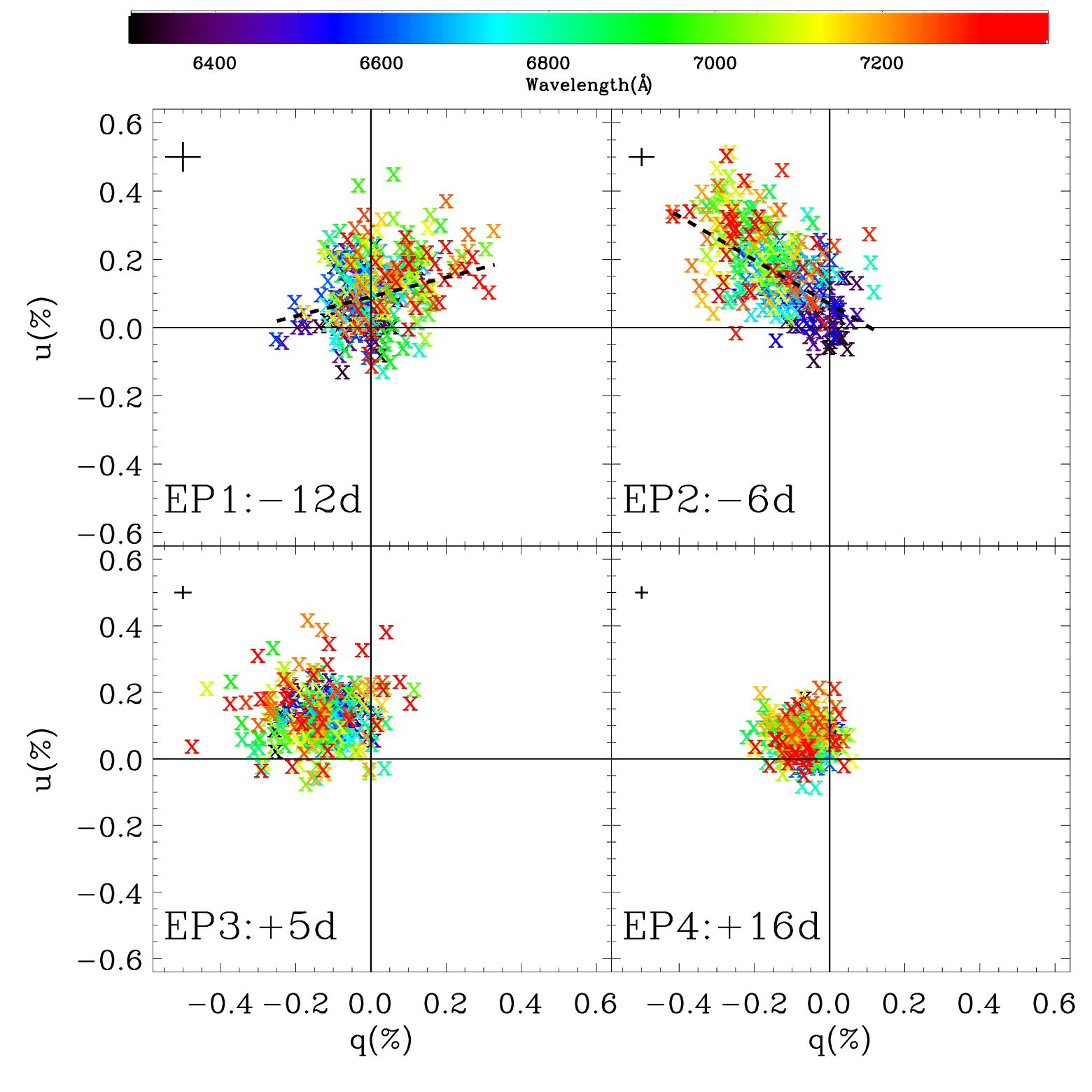}
\caption{The red continuum, 6300 - 7400\AA, in the $q$/$u$ plane, 
for Epochs 1-4. There is an indication of axial symmetry in the Epoch 2 
polarization, pointing in a different direction than seen in Epoch 1. 
Dashed lines show linear fits to the data points. The Epoch 4 data 
is not aligned with the origin, likely due to using a constant rather than 
an actual Serkowski function when performing ISP subtraction.} 
\label{red_cont}
\end{figure*}

\subsection{Evolution of Continuum Polarization} 

The red wavelengths of the continuum emission exhibit  
polarization reaching up to $\sim$0.4\% in Epoch 2. 
The wealth of line polarization features in SN~2011fe makes it 
difficult to measure the continuum polarization at shorter wavelengths, 
but there is the suggestion of polarization at the $\sim$0.1\% level 
between Features $C$ an $A$ in Epochs 1-3 (Figs. 4-6).  A similar 
level of red continuum emission was reported for 
SNe~1999by (Howell et al. 2001), 2001el (Wang et al. 2003), 
2004S (Chornock \& Filippenko 2008), and 2005ke (Patat et al. 2012). 
In Figure \ref{vs_99by_05ke}, we show red-continuum polarization 
of SN~2011fe during Epoch 2 compared to SNe~1999by and 2005ke, with 
2011fe providing a reasonable match with 2005ke. 
The increase in polarization toward longer wavelengths has been proposed 
as beind due to an overall oblate geometry combined with the 
decreased importance at red-wavelengths of line scattering depolarizing 
the observed light (Wang 1997). This hypothesis has been theoretically supported 
(e.g. Patat et al. 2012; Bulla et al. 2015), and suggests that longer wavelengths 
are the best for studying the overall degree of ashpericity in the SN ejecta. 
The continuum polarization in the 6300 -- 7400 \AA\ wavelength range 
varied with epoch, perhaps best presented in the form of a $q/u$ scatter 
plot (Figure \ref{red_cont}). The continuum polarization is evident during 
Epochs 1 \& 2, but is absent by Epoch 4. Linear fits to the Epochs 1 \& 2 data 
suggest a change in angle of $\sim$20 degrees, although the linearity 
of the scatter is only apparent in Epoch 2. The 
Bulla et al. (2015) oblate geometry model features the polarization of the 
red continuum dropping between 15 post-exposion to 25 days post-explosion. 
The time variability of the continuum polarization seen in SN~2011fe might 
be evidence of an oblate geometry. It would be interesting to determine 
the level of oblateness that best describes SN~2011fe and then revisiting the 
Patat et al. (2013) study of the NaD variability, but with an oblate, rather 
than spherical emitting region.

\section{Comparisons with Polarization in other SNe~Ia}

\subsection{Line Features $A$ - $D$}

The earliest line polarization feature, $A$, is most polarized 
during Epoch 1, and is gone by the third epoch. 
SN~2012fr, 2002bo and SN~2006X are the only other normal SN~Ia with published 
observations at similar epochs.
Maund et al. (2013) report no line features in that wavelength range for 
SN~2012fr, indeed they use the 5100 -- 5300 \AA\ wavelength range as an 
``intrinsically depolarized" region. Similarly, no equivalent line features  
are apparent for SN~2002bo in Figure 1 of Wang et al. (2007). A daily 
sequence of VLT spectropolarimetry of SN~2006X does show a polarized 
feature in that wavelength range (referred to as $\lambda$5051 \AA\ in that work), 
with the polarization reaching 0.37\% at -6 days. 
SN 2006X's feature was narrower and did not show a well-defined peak as in SN~2011fe.                     
SNe~2002bo \& 2006X are of the HVG/HV group, so 
comparisons with 2011fe are perhaps less appropriate than comparisons 
with SN~2012fr, a member of the LVG/LV group, as the HV/LV differentiation is based on 
the pre-peak epochs. However, the presence/absence 
of feature $A$ in early-epoch spectropolarimetry 
does not appear to be strictly correlated with the HV/LV categorization. 
It is worth mentioning that, SN~2012fr is a NUV-red event without a 
detection of unburned carbon (see Brown et al. 2014 [SOUSA] for UVOT 
photometry and Childress et al. 2013 for a spectral study), whereas 
SN~2011fe is a NUV-blue event with a detection of unburned carbon 
(Milne et al. 2013; Pereira et al. 2013). 
A handful of supernovae have also shown line polarization near 4800 \AA, but at 
post-maximum epochs (+9 days and later for SNe 1997dt, 2003du, and 2004S), causing the 
feature to be identified as Fe ~{\sc ii} (Leonard et al. 2005).  
Each of these supernovae only had a single epoch of observation, so we cannot say if the 
line polarization was observable at earlier epochs.

\begin{figure*}
\epsscale{0.70} 
\plotone{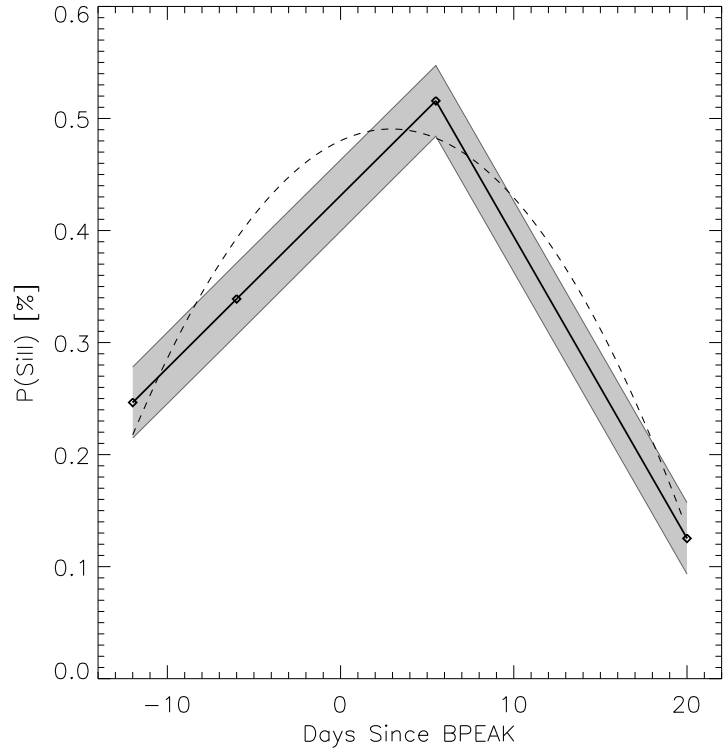}
\caption{The time evolution of the polarization of the 
Si~{\sc ii}$\lambda$6355 \AA\ absorption feature. Whereas some SNe~Ia studied 
in Wang et al. (2007) were found to reach the largest Si~{\sc ii} polarization roughly 
5 days before optical maximum, SN~2011fe clearly reaches the largest polarization 
post-optical maximum, if assuming a parabolic fit to the data with a shape similar to 
Wang et al. (2007) (dashed line). Error bars were generated by determining the standard 
deviation of P$_{SiII}$ estimations when $q$(ISP) and $u$(ISP) change according to 
the averages from Epochs 4-9, as shown in Figure 2. A continuum polarization level of 
$\sim$0.08\% was not subtracted from P$_{SiII}$.} 
\label{fig:psi2_vs_time}
\end{figure*}

Historically, the peak of the Si~{\sc ii} $\lambda$6355 \AA\ line, Feature $B$, has been 
used as the measure of P$_{SiII}$. Although this introduces a dependence on 
spectral resolution which would be improved through the use of equivalent widths, 
for the sake of comparison with other SNe~Ia in the literature, in this work we 
also use this method.  
As shown in Figure \ref{fig:psi2_vs_time}, the evolution of the peak polarization of the 
$\lambda$6355 \AA\ line is 0.24\%, 0.34\%, 0.52\% for Epochs 1-3, respectively. 
In that figure, we generate error bars by determining P$_{SiII}$ for each point 
in a grid of estimates when $q$(ISP) and $u$(ISP) are each varied by $\pm$0.05\%, reflecting 
the variation of the $q$(ISP) and $u$(ISP) estimates shown in Figure 2. 
The error bars are then the standard deviation of the 
estimates in that grid, and at a level of $\sim$0.03\%, 
they show that the ISP estimation does not 
appreciably affect the evolution of P$_{SiII}$. 
We keep with the formalism of P$_{SiII}$ estimation in the literature and do 
not subtract the continuum polarization (which we estimate to be $\sim$0.08\%). 
Thus, by Epoch 4, the polarization is consistent with zero. 
This evolution is surprising, as earlier works presented SNe~Ia for which Feature $B$ 
clearly peaks before the optical maximum (Wang et al. 2007; Patat edt al. 2009). However, 
our time sequence clearly favors this feature reaching a maximum after the optical
peak, which is fundamentally different than the scenario presented in those previous 
works. Porter et al. (2016) reports that SN~2014J also features a
late-peaking P$_{SiII}$, so we do not consider SN~2011fe to be
anomalous.

Concentrating on Feature $C$, 
Mg~{\sc ii} $\lambda4471$ \AA ~was polarized in SN 2006X where the line polarization 
extended over a narrower range than SN~2011fe's Feature $C$ between 4150-4300 \AA (Wang et al. 2007).  
The authors mention the feature is significant only at -3 and -1 days when it peaked 
at 0.5\%, but was present starting as early as -8 days.  Mg~{\sc ii} was also polarized 
in SN 2004dt at -7 days between 4100-4300 \AA ~and reached a peak of 
$\sim 1\%$ polarized (Wang et al. 2006). 

Feature $D$ is mentioned as being due to S~{\sc ii} for SNe~2005ke (Patat et al. 2012) and 
2012fr (Maund et al. 2013), but not in SN~2004S (Chornock \& Filippenko 2008). 

In terms of the relative evolution of the features, for SN~2011fe, Feature $B$ increases in 
polarization as Feature $A$ decreases, whereas in SN~2006X, Feature $A$ is always weaker than 
Feature $B$.  

To summarize, although most of the attention of SNe~Ia polarization literature has focused 
on the Si~{\sc ii} $\lambda6355$ \AA ~and Ca II lines, Features $A$, $C$, and $D$ as identified 
in SN~2011fe are present in other fairly well observed SNe~Ia with polarization measurements.  
Specifically, Mg~{\sc ii} has been correlated with Feature $C$, S~{\sc ii} with Feature $D$, 
and Feature $A$ has been matched with Si~{\sc ii} at pre-maximum epochs and Fe~{\sc ii} post-maximum.

\subsection{Previously-reported Correlations}

%The evolution of the $\sim\lambda$5050\AA\ line polarization feature through 
%Epochs 1-3 is thus 0.30\%, 0.25\%, 0.11\%, while for the $\lambda$6355\AA\ 
%line it is 0.20\%, 0.32\%, 0.52\%, respectively, after subtracting the 
%continuum polarization.  

SN~2011fe can be included with other SNe~Ia in studies of the time evolution of 
Si~{\sc ii} $\lambda$6355~\AA\ polarization. 
Wang et al. (2007) report a postive correlation between the width of the optical 
peak and the strength of the Si~{\sc ii} $\lambda$6355 \AA\ line polarization 
at -5 days (P$_{SiII}$). For SN~2011fe, Munari et al. (2013) report 
$\Delta m_{15}(B)$ = 1.108, Richmond \& Smith (2012) report 
$\Delta m_{15}(B)$ = 1.2, and Pereira et al. (2013) report 
$\Delta m_{15}(B)$ = 1.103. Using a rough average of those values, 
$\Delta m_{15}(B)$ = 1.15, the 
Wang et al. (2007) formula predicts P$_{SiII}$ = 0.55\%. This is consistent with 
our maximum measured polarization, 0.52\% at +5 days. Since Wang et al. (2007) chose -5 days 
as their standard epoch, and we found SN~2011fe peaks post-optical peak, we will also 
consider equally the -6 day polarization, which we measure to be 0.34\%. 
Because the temporal sampling of Si~{\sc ii} $\lambda$6355 \AA\ polarization 
is so poor for most SNe~Ia, for comparisons with other SNe, we consider 
SN~2011fe to be between 0.34\% and 0.52\%, the -5 day and highest measured values, 
respectively. In the upper left panel of Figure \ref{fig:psi2_vs_others},
we show these two choices of P$_{SiII}$ plotted versus $\Delta m_{15}(B)$ with other SNe~Ia, 
concluding that SN~2011fe falls within the scatter of the relation for 
either choice of P$_{SiII}$. 

SN~2011fe also falls within the scatter of other 
SNe~Ia for P$_{SiII}$ plotted versus the equivalent width of Si~{\sc ii} $\lambda$6355 \AA\. 
That relation tests the strength of polarization as being only due to the line 
strength. In the lower left panel of Figure \ref{fig:psi2_vs_others}, we treat the    
Maund et al. (2010) suggestion of a linear correlation between the Si~{\sc ii} 
velocity gradient (e.g. $\dot{v}_{SiII}$: HVG/LVG) and P$_{SiII}$. 
Pereira et al. (2013) report 
$\dot{v}_{SiII}$ = 59.6 $\pm$ 3.2 km s$^{-1}$ d$^{-1}$, leading to a predicted 
P$_{SiII}$ = 0.62\%. This is again slightly larger than our measured value of 
0.52\%, as seen in the lower left panel of Figure \ref{fig:psi2_vs_others}, but 
SN~2011fe falls comfortably within the scatter of the other SNe~Ia 
plotted.  

\begin{figure*}
\epsscale{0.80}
\plotone{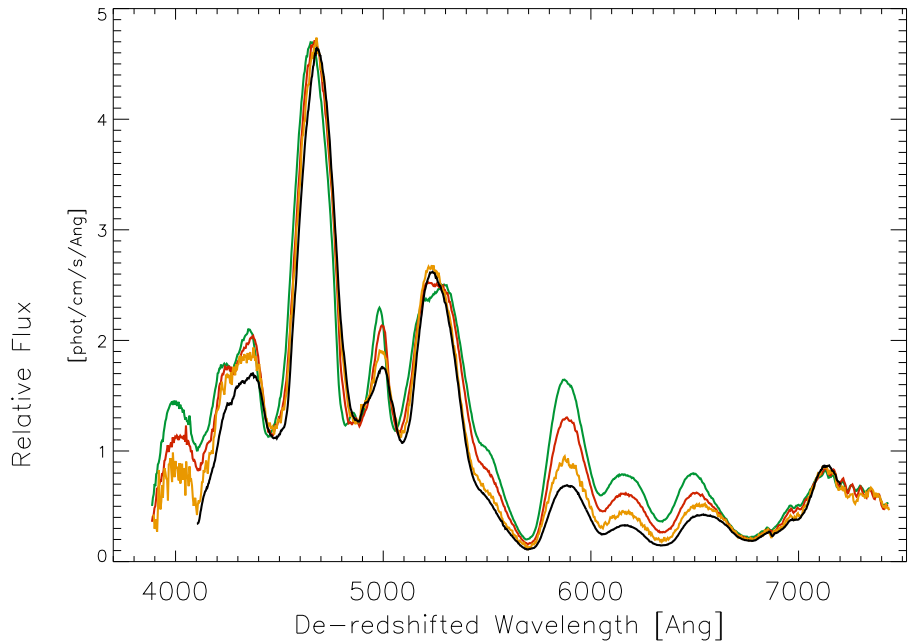}
\caption{Spectral sequence of the nebular flux spectra of SN~2011fe. 
Epochs 7-10 show the evolution of forbidden [Fe] and [Co] lines. 
Epochs 7-10 are colored as 7(green), 8(red), 
9(orange), 10(black), are normalized to the $\sim$4600 \AA\ line. 
The progression of the [CoIII] complex in the 5700 -- 6700 \AA\ wavelength 
range is apparent.}
\label{fig:late-spec-seq}
\end{figure*}

\begin{figure*}
\epsscale{0.90}
\plotone{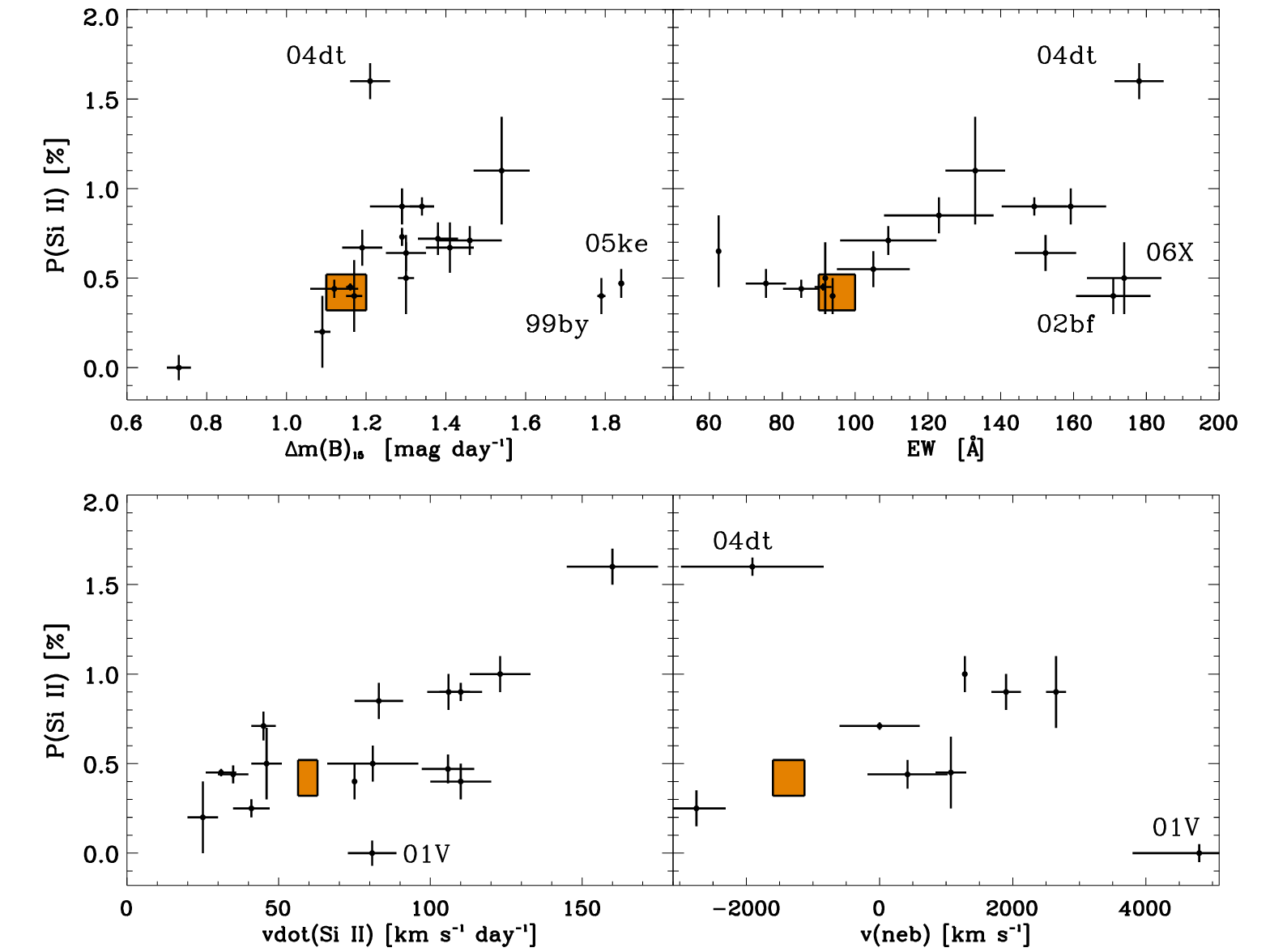}
\caption{Polarization of the Si~{\sc ii} $\lambda$6355 \AA\ absorption feature 
compared with potentially correlated parameters. Wang et al. (2007) claim a 
correlation between P$_{SiII}$ and peak width (upper left panel). It has been 
suggested that there is a correlation between P$_{SiII}$ and the equivalent 
width of the Si~{\sc ii} $\lambda$6355 \AA\ absorption feature (upper right panel). 
Maund et al. (2010) claim a correlation between P$_{SiII}$ and the velocity gradient of the 
Si~{\sc ii} $\lambda$6355 \AA\ absorption feature (lower left panel). 
Maund et al. (2010) also investigated claims of a correlation between P$_{SiII}$ 
and the nebular velocity of the emission feature, [FeII] 7155 \AA\ (lower right panel). 
The filled boxes show our measured values for SN~2011fe. The range in polarization 
reflect the difference between the -5 day polarization and the maximum 
measured polarization. The range in peak width and $\dot{v}_{SiII}$ are 
from the literature. The range in v$_{NEB}$ is from the minimum and maximum 
measured in our Epoch 5,6,7 nebular spectra. Non-2011fe data are from Wang et al. (2007), 
Periera et al. (2013), Silverman et al. (2012), Hachinger et al. (2006) and    
Maund et al. (2010), and references therein.}
\label{fig:psi2_vs_others}
\end{figure*}

Maund et al. (2010) also investigated a linear correlation between the strength 
of the -5 day Si~{\sc ii} $\lambda$6355 \AA\ line polarization and the nebular 
phase velocity of Fe-group elements, based upon a claimed correlation between 
the nebular velocity and the velocity gradient of the Si~{\sc ii} $\lambda$6355 \AA\  
absorption feature by Maeda et al. (2010). Maund et al. (2010) specifically 
measured the shifting of the [FeII] $\lambda$7155 \AA\ and [NiII] $\lambda$7378 \AA\  
emission lines. The wavelength range of our spectra (Fig. \ref{fig:late-spec-seq}) 
only permit the study of the [FeII] $\lambda$7155 \AA\ line; 
we find shifts of -1128, -1149 and -1601 km s$^{-1}$ 
for Epochs 5,6 and 7, respectively. The lower panel of Figure \ref{fig:psi2_vs_others} 
shows that SN~2011fe is similar to a number of other SNe~Ia which have a 
measurement of the [FeII] $\lambda$7155 \AA\ shift, but as noted by 
Maund et al. (2010), there is no clear correlation between these parameters. 
SNe 2001V and 2004dt are identified because each is peculiar and not considered 
members of the ``normal" group.  

Despite peaking at a later epoch than the Wang et al. (2007) standard 
epoch, the Si~{\sc ii} $\lambda$6355 \AA\ polarization of SN~2011fe 
falls to zero by 
+16 days. Spectra of normal SNe~Ia at a similar epoch were presented 
for SNe~1997dt (+21d) and 2003du (+18d) by Leonard et al. (2005), 
for SN~2002bo (+14d) by Wang et al. (2007) and 
for SN~2012fr (+24d) by Maund et al. (2013). These comparison 
spectra all show some evidence for Si~{\sc ii} $\lambda$6355 \AA\  
polarization, although for SN~2012fr the feature is quite weak. 
Porter et al. (2016) further investigates the time evolution of the 
Si~{\sc ii} $\lambda$6355 \AA\ polarization for a collection of SNe~Ia. 

\section{SUMMARY}

By virtue of being a very nearby SN~Ia that was discovered at an 
extremely early epoch, SN~2011fe has proven to be one of the best studied 
SNe~Ia of all time. Spectropolarimetry of this SN obtained over 
4 epochs has revealed both line and continuum polarization, with both 
components exhibiting time-variability. 

Utilizing the $SYNAPPS$ algorithm, we present 3 possible interpretations 
for the line polarization features in the spectra from the initial epoch 
to later epochs with no line polarization. Interestingly, the maximum polarization 
of the common Si~{\sc ii} $\lambda$6355 \AA\ feature ($B$) 
happens at a later epoch than has been seen for other SNe~Ia. 

These polarization features add to the study of this well-observed 
normal, NUV-blue, LV, unburned-carbon bearing SN~Ia. Over the course 
of a multi-year campaign, we plan to observe enough SNe~Ia to 
search for similarities and differences between individual events 
as a function of known parameters, exploring which characteristics 
of a SN~Ia explosion drive polarization. 

\acknowledgements
\footnotesize

This work was partially supported by NSF Collaborative Research 
Award \#AST-1210599. 
P.A.M.\ acknowledges support from NASA ADP grant NNX10AD58G.
P.S.S.\ acknowledges support from NASA/{\it Fermi} Guest Investigator
grant NNX09AU10G. N.S. received partial support from NSF graqnt AST-1515559.
B.T.J. acknowledges support from the NSF, through its funding
of NOAO, which is operated by AURA, Inc., under a cooperative
agreement with the NSF. Some observations reported here were obtained at the 
MMT Observatory, a joint facility of the University of Arizona and the 
Smithsonian Institution.

\end{document}